\documentclass{article}

\usepackage{RR}
\RRNo{6471}
\usepackage{graphicx}
\graphicspath{{figures/}}
\usepackage{cite}
\usepackage{amsmath}
\usepackage{amsfonts}
\usepackage{mathrsfs}
\usepackage[latin1]{inputenc}
\usepackage[T1]{fontenc}

\RRtitle{WiFly: expérimentation avec un Réseaux de capteurs et des coordonnées virtuelles}
\RRetitle{WiFly: experimenting with Wireless Sensor Networks and Virtual coordinates\thanks{This work was partially supported by the French Ministry of Research under contract ARESA ANR-05-RNRT-01703.}}
\titlehead{WiFly: WSNs and Virtual coordinates}

\RRauthor{
   Thomas Watteyne\thanks[ft]{France Telecom R\&D, Meylan, France. firstname.lastname@orange-ftgroup.com}\thanks[citi]{ARES INRIA / CITI, INSA-Lyon, F-69621, France. firstname.lastname@insa-lyon.fr}
\and
   Dominique Barthel\thanksref{ft}
\and
   Mischa Dohler\thanks[cttc]{Centre Tecnològic de Telecomunicacions de Catalunya (CTTC), Barelona, Spain. mischa.dohler@cttc.es}
\and
   Isabelle Augé-Blum\thanksref{citi}
}

\authorhead{Watteyne, Barthel, Dohler, Augé-Blum}

\RRdate{January 2008}

\RRresume{L'étude expérimentale est importante lorsque l'on crée des protocoles de communication pour réseaux de capteurs sans fils. Les couche protocolaires bas niveau ont un grand impact sur les performances des couches supérieures, et la complexité des phénomènes ne peut pas être entièrement capturé par l'analyse mathématique ou la simulation. Dans ce rapport, nous décrivons le processus complet, depuis la mise en place d'une architecture de communication efficace en énergie et auto-organisante (couches MAC, routage et application), jusqu'au déploiement réel.

L'architecture de communication présentée comporte un protocole MAC qui évite la construction et la maintien de tables de voisinage, ainsi qu'un protocole de routage inspiré par les protocoles de routage géographique. Celui-ci s'appuie sur des coordonnées virtuelles des noeuds, indépendantes de leurs coordonnées réelles. L'application consiste en un noeud de collecte mobile interrogeant un réseau de capteurs sans fil, à partir de requêtes émises par une station de base déconnectée du réseau.

Après le processus de mise en place de cette architecture, nous vérifions son bon fonctionnement par simulation, et nous effectuons une étude temporelle. Cette dernière est utile pour calculer la vitesse maximale du noeud de collecte mobile. Nous détaillons les phases de l'implémentation, et les résultats des expérimentations préliminaires (consommation énergétique à la couche PHY, probabilité de collision au niveau MAC, et routage). Finalement, nous présentons l'expérimentation finale où nous montons le noeud de collecte mobile sur un avion radioguidé.}

\RRmotcle{Réseaux de capteurs, expérimentation, communication sans fil multi-sauts, coordonnées virtuelles, noeud de collecte mobile.}

\RRabstract{Experimentation is important when designing communication protocols for Wireless Sensor Networks. Lower-layers have a major impact on upper-layer performance, and the complexity of the phenomena can not be entirely captured by analysis or simulation. In this report, we go through the complete process, from designing an energy-efficient self-organizing communication architecture (MAC, routing and application layers) to real-life experimentation roll-outs.

The presented communication architecture includes a MAC protocol which avoids building and maintaining neighborhood tables, and a geographically-inspired routing protocol over virtual coordinates. The application consists of a mobile sink interrogating a wireless sensor network based on the requests issued by a disconnected base station.

After the design process of this architecture, we verify it functions correctly by simulation, and we perform a temporal verification. This study is needed to calculate the maximum speed the mobile sink can take. We detail the implementation, and the results of the off-site experimentation (energy consumption at PHY layer, collision probability at MAC layer, and routing). Finally, we report on the real-world deployment where we have mounted the mobile sink node on a radio-controlled airplane.}

\RRkeyword{Wireless Sensor Networks, experimentation, multi-hop wireless communication, virtual coordinates, mobile sink node.}

\RRprojet{ARES}
\RRtheme{\THCom}
\URRhoneAlpes

\begin{document}

\makeRR


\section{Introduction and related work}

Much effort has been put during the last 5-10 year into Research on Wireless Sensor Networks (WSNs). Numerous conferences, journals and special issues are dedicated to these networks, and new solution appear on a weekly basis. Despite all this activity, a surprisingly low number of actual deployment examples have been made public. Whereas rolling out a solution can be considered more part of engineering rather than Research, we argue that physical implementation confronts the researcher with important on-field constraints. As solutions for WSNs are cross-layered, and as these solutions are largely impacted by lower layers (e.g. wireless transmission), real world confrontation has a very beneficial impact on Research.

Real-world deployment has been largely simplified by the appearance of commercial products. The most-known MICA wireless sensor nodes have been developed by laboratories at University of California at Berkeley. They were initially commercialized by Crossbow (Mica2 in 2002, Mica2dot in 2003), the latest versions (Tmote SKY in 2004 \cite{polastre05telos}, Tmote Mini in 2007) are brought to the market by Moteiv, a spin-off company of the University of California at Berkeley. On-going Research is aiming at developing energy-harvesting nodes which collect data from their environment, radically changing the energy-constrained assumption made for WSNs. In \cite{jiang05perpetual}, the authors for example attach a solar panel to an early version of the Tmote SKY nodes.

A pioneering team at Berkeley lead the smart dust project, which used the early versions of the Mica2 motes to do proof-of-concept demonstration. An early experiment in 2001 involved a autonomous radio-controlled airplane which dropped sensors along a highway to monitor the passing of large military vehicles. The plane continuously passed above them to collect the measured data, which was then transfered back to a base station.

In 2002, a 43-node network was deployed on an uninhabited island 15km off the coast of Maine, USA\cite{szewczyk04lessons}. This network was used to monitor the migration and nesting habits of birds. With the monitored data being available online in real time, this deployment can be seen as a milestone and an early public demonstration of WSNs. The same team deployed a network to monitor trees in a tropical forest\cite{culler04overview}. In \cite{langendoen06murphy}, the authors didactically describe the numerous problems one can face during real-world deployments of WSNs.

Since 2005, companies have been emerging which provide services entirely based on WSNs. One interesting example is Coronis, a French start-up company, specialized in automated meter reading. It's first big deployment involved a network of 25,000 nodes attached to the home water meters of a medium-sized city\cite{dugas05configuring}. It now has sold over a million of those sensors worldwide.

The company Arch Rock received major attention lately\cite{clark06sensor}. It commercializes an off-the-shelf solution for small to medium scale monitoring WSNs (typically less than 100 nodes). Its current solution involves packaged Tmote SKY nodes which communicate with a small personal computer as sink node. This computer in turn is connected to the Internet and with the use of Web Services allows the integration of this network into larger applications.

The main contributions of this work are:
\begin{itemize}
   \item we present a \textbf{complete energy-efficient self-organizing communication architecture} for Wireless Sensor Networks. This solution combines MAC and routing protocols into a cross-layered solution, and is particularly suited for low-throughput, dynamic and energy-constrained applications.
   \item we implement this solution on a medium-sized network. A mobile sink consisting of a radio controlled airplane is used to interrogate the WSN based on requests issued by a remote base station.
\end{itemize}

Whereas our implementation resembles the early implementation done by the Smart Dust team, the key difference is that the mobile sink communicates with a complete WSN and not a series of individual nodes. In the latter case, the networking problems were largely simplified as the multi-hop nature of node-to-sink communication was essentially removed. Having a real multi-hop WSN raises interesting problems such as self-organization and real-time communication.

The remainder of this report is organized as follows. In Section~\ref{architecture}, we describe the communication architecture used in the WSN. The experimental setup is presented in Section~\ref{setup} together with hardware details and frame durations. Section~\ref{real-time} focuses on real-time communication, and calculates the maximum speed the mobile sink may move at. Simulations results are presented in Section~\ref{simulation}. Experimental results are split in two sections. Section~\ref{off-site experimental} focuses on preliminary experiments conducted off-site, Section~\ref{on-site experimental} presents the results obtained during deployment. This report is concluded, and future work is presented in Section~\ref{conclusion}.


\section{The communication architecture}
\label{architecture}


\subsection{Overview}

In this work, we aim at evaluating the performance of the stack represented in Table~\ref{table:stack}, which combines the 1-hopMAC medium access control protocol\cite{watteyne061hopmac}, the 3rule routing protocol\cite{watteyne07geographic} and the use of virtual coordinates\cite{watteyne07using}. The main challenge is to form a \textbf{complete energy-efficient self-organizing communication architecture} from these protocols. This includes adapting the different layers one to the other. In the subsequent subsections, we detail the adaptations which were needed.

\begin{table}
\begin{center}
\begin{tabular}{|l|}
\hline
\textbf{Application}\\
connectivity graph construction\\
\hline
\hline
\textbf{Routing}\\
3rule routing\\
virtual coordinates\\
\hline
\hline
\textbf{Medium access control}\\
1-hopMAC\\
\hline
\hline
\textbf{Physical layer}\\
EM2420 module\\
\hline
\end{tabular}
\caption{The communication stack}
\label{table:stack}
\end{center}
\end{table}


\subsection{Adapting the 1-hopMAC protocol}

1-hopMAC\cite{watteyne061hopmac} is a medium access control protocol for WSNs which avoids the need to maintain a neighborhood list. Maintaining such a list at each node would mean periodically exchanging Hello packets. This can turn out to be very energy consuming, as Hello packets need to be exchanged even when the network sits idle. In a forest fire detection scenario (or any scenario with low throughput), the network would deplete its energy by periodically exchanging Hello packets whereas there is no useful data to transmit.

1-hopMAC tackles this problem in a fully on-demand solution. When a node wants to send some data, it issues a request to which all of its neighbors answer using a backoff timer inversely proportional to some metric. The node which answers first is elected relaying node. The metric attached to each node does not need to be unique, but it should be carefully chosen by the routing layer so that following a path of decreasing metric leads to the sink node. To be fully energy-efficient, 1-hopMAC uses the variant of preamble sampling described in \cite{bachir06micro}, which enables idle duty cycle of as low as 1\%. The idle duty cycle accounts for the percentage of time a node has its radio on ("duty cycle") when no information is sent or received ("idle").

In the original 1-hopMAC protocol, the sending node could take action after receiving the first acknowledgment message. As our routing protocol (described next) needs the complete list of neighbors, we modify the 1-hopMAC protocol by asking the sending node to wait for all the ACK messages before taking action. This mode is called the basic mode in \cite{watteyne061hopmac}.

Note that, as such, if nodes have very close metric, the ACK messages could be separated by a duration which is so small that ACK messages would collide. We address and answer this problem in subsection~\ref{collision}.


\subsection{Adapting the 3rule routing protocol}
\label{adapting_3rule}

It has been shown in \cite{watteyne07geographic} that current geographic routing protocols such as GFG\cite{frey06delivery} or GPSR\cite{karp00gpsr} suffer from inaccurate positioning systems. Inaccurate position can even cause those routing protocols to fail, \textit{i.e.} they do not deliver their message although there exists a physical path. The 3rule routing protocol (presented and called LeftHandGeoPR in \cite{watteyne07geographic}) asks each node to append its identifier to the packet header. With this information, it efficiently chooses the next hop node using a distributed version of the well known depth first search algorithm in a tree. Although the 3rule protocol increases the size of the packet header, it is shown that it achieves a 100\% delivery ratio independently from the positioning accuracy, with a hop count identical to GFF or GPSR.

Its robustness lead us to chose this routing protocol for our communication architecture. Nevertheless, the presence of a mobile sink somewhat complicates the problem as the sink may have moved when the message reaches its original destination. When this happens, we restart the 3rule routing protocol by erasing the sequence of traversed nodes in the header. We show by simulation in Section~\ref{simulation} that the protocol restarts only a limited number of times, and that this number quickly decreases with lower speeds of the sink or number of neighbors increasing.

Yet, using a geographical-based routing protocol implies that nodes know their positions which is a costly assumption (both in terms of money and energy). As GPS-like solutions can not be count on, we recently showed in \cite{watteyne07using} that virtual coordinates can overcome this problem. By iteratively applying centroid transformation to initially random coordinates, the number of hops using the 3rule routing protocol on those virtual coordinates drops sharply. It is shown that with about 10 centroid rounds, the number of hops drops by more than 50\% compared to the fully random case. Current work shows that with another type of virtual coordinate update, the network converges to a state where path length only exceeds the shortest path by 4\%. Details will be given in subsequent publications.

The use of virtual coordinates is particularly suited for the case of a mobile sink. Indeed, as in our solution the sink keeps the same predefined virtual position (regardless of its real physical position), it does not need to periodically inform the other nodes of its position. This, we believe, is a major advantage of using virtual coordinates, and is much simpler than the classical periodic heartbeat\cite{shim06locators} or rendez-vous point\cite{shin05railroad} solutions.


\section{Experimental setup and implementation details}
\label{setup}


\subsection{The experimentation framework}

\begin{figure}
\centering
\includegraphics[width=0.50\textwidth]{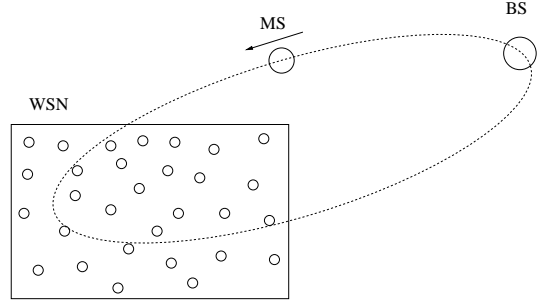}
\caption{The experimental framework setting.}
\label{figure:setting}
\end{figure}

In order to test the communication architecture presented in Section~\ref{architecture}, we need to add some protocols to operate in a realistic environment with a real application. Our target application is an on-demand tracking system, where we assume only one node answers a specific request. A mobile sink is given a query by a base-station which is disconnected from the network (Phase 1). It travels to the WSN network where it communicates this query to a random node (Phase 2). The query is then broadcasted in the network (Phase 3). The node which holds the answer (called source node) transmits it using our communication architecture to the mobile sink (Phase 4). The mobile sink acknowledges this reception (Phase 5), travels back to the base station to which it transmits the data (Phase 6). Refer to Fig.~\ref{figure:setting} for an illustration of the experimentatal framework.

From the previous description, it is clear that our main interest is how the data is transmitted from the source node to the mobile sink, \textit{i.e.} Phase 4. All other phases are used to provide a real-world evaluation framework, but are not the core of our study. That's why these phases may use simplistic/suboptimal solutions.

We now detail the different phases, introducing the packet names:
\begin{itemize}
   \item \textbf{Phase 1: Data Request.} This phase involves the base station ($BS$) and the mobile sink ($MS$). The $BS$ periodically sends Data Request messages with period $T_{DRp}$ and of duration $D_{DRp}$. These $DRp$ messages are formed by a sequence of micro-frames, which each contain the number $Seq$ of remaining micro-frame in the $DRp$ message. The $MS$ checks whether the medium is free every $T_{cca}$ and during $D_{cca}$. $D_{cca}$ is chosen such that it hears a complete micro-frame when a sequence of micro-frames is sent ($D_{cca} \geq T_{DRp}+D_{DRp}$). When the $MS$ correctly receives a micro-frame, it calculates using $Seq$ when the $BS$ finishes to send all the microframes the $DRp$ message consists of, and sends an $ACK$. Whenever the $BS$ is not sending a $DRp$ message, it is listening to medium, waiting for the $ACK$ message. Once the $ACK$ is sent, the $MS$ enters phase 2.
   \item \textbf{Phase 2: starting Broadcast Request.} This phase involves the mobile sink ($MS$) and the WSN. After phase 1, the $MS$ periodically sends Broadcast Requests with period $T_{BRp}$ and of duration $D_{BRp}$. Similar to the $DRp$ messages, these $BRp$ messages are formed by a sequence of micro-frames. When a node hears a $BRp$ micro-frame, it waits for the end of the $BRp$ and starts a random backoff $B_{BR}$ uniformly chosen within a contention window of length $W_{BR}$. During its backoff duration $B_{BR}$, it listens for other potential $BRp$ message. If it receives a second $BRp$, it cancels $B_{BR}$ and restarts it $D_{BRp}$ later. When $B_{BR}$ elapses, the node sends a $BRp$. Upon hearing a relayed $BRp$, the mobile sink knows its $BRp$ has been heard, and it enters directly Phase 5.
   \item \textbf{Phase 3: Broadcast Request.} This phase only involves the WSN. The backoff-based algorithm described in Phase 2 is carried out between all nodes. Its goal is to flood the complete network. Note that each node will send exactly one copy of $BRp$. The node which holds the answer to the request identifies itself. Upon receiving the $BRp$, it does not start the $B_{BR}$ backoff but rather the $B_{SRC}$. This backoff is used to wait for the flood to pass. Upon elapsing $B_{SRC}$, the source node enters Phase 4. Note that all other nodes enter Phase 4 after receiving a second copy of the $BRp$, or after relaying it.
   \item \textbf{Phase 4: Routing.} This phase only involves the WSN. This is the phase we are interested in. The source node sends a message to the $MS$ using the communication architecture described in Section~\ref{architecture}. We call $DATA$ the data messages, and $B_{ACK}$ the backoff taken within a contention window $W_{RR}$. Note that $B_{ACK}$ is proportional to the metric of the node, which is the virtual distance to the $MS$. In order to be more robust, we ask each node to listen to the medium for a fixed duration $B_{RR}$. If during this period, it does not hear another node retransmitting $RRp$, it assumes it was lost and retransmits it.
   \item \textbf{Phase 5: DATA reception by mobile sink.} This phase involves the mobile sink ($MS$) and the WSN. The $MS$ has entered this phase after Phase 2, and is waiting for the $DATA$ to reach it. It runs the 1-hopMAC protocol described in Section~\ref{architecture} but has a metric of 0. After receiving the $DATA$, it sends an $ACK$ to inform the sender not to retransmit the message. The $MS$ then switches to Phase 6. All nodes, after successfully relaying the message switch to Phase 2.
   \item \textbf{Phase 6: DATA retrieval by base station.} This phase involves the base station ($BS$) and the mobile sink ($MS$). Recall that the $BS$ periodically sends $DRp$ messages. Phase 6 is similar to Phase 1, the only difference being that the $MS$ answers to the $DRp$ with a $DATA$ packet.
\end{itemize}


\subsection{Parameters and hardware}

For experimental testing of our communication architecture, we have used a WSN composed of 20 Ember EM2420 nodes. The core components of these sensors are a Ember/Chipcon CC2420 radio chip, and a Atmel AtMega128 micro-controller. Some nodes were equipped with sensing devices, push/slide button and light meters. All nodes were programmed using a component based language called Think, which is developed at France Telecom R\&D. Unlike the TinyOS or Contiki operating systems, Think is based on a set of components which are compiled together to form a binary code, which is then loaded onto the nodes. This components approach offers great flexibility and code re-use as individual components such as the scheduler or a specific routing protocol do not need to be reprogrammed when changing application. This is also true for changing platform, which enabled us to use two platforms.

As those nodes are very constrained, we were limited by the following. The transmission queue is limited to 128 bytes, which is thus the maximum size of the $DATA$ packets. The EM2420 module is 802.15.4 enabled, but while we completely replaced its MAC protocol with 1-hopMAC, we were still bound by the hardware to use 2 byte addresses. The nodes communicate at 250 kbps, with one physical symbol encoding 4 bits of data. As for our simulations, we assumed having a 25 node network, with an average number of neighbors of 5. The EM2420 needs $T_{RxTx}=192 µs$ to switch between reception and transmission states, which we needed to take into account during implementation of our protocols.

The base station is formed by a laptop connected via a RS232 link to the EM2420 development kit. This connection is only used to monitor the activity of the base station, which is really the node connected to the development kit. As we wanted to test a large range of mobile sink speeds, we have used an MS 2001 radio-controlled airplane with an EM2420 node attached to it.

The implemented application is the following. We want to determine the network topology, \textit{i.e.} which are the neighbors of each node. For this, the $BS$ asks for the list of neighbors of a specific node, by putting the node identifier in its $DRp$, as specified in the nect subsection. Each plane rotation will enable the base $BS$ to learn the neighbors of a specific node, and after a series of rotations, the $BS$ will be able to construct the connectivity graph of the network. Note that this is just a proof-of-concept application, and this connectivity graph is not used by the MAC and routing layers.


\subsection{Packet format, sizes and durations}

In Table~\ref{table:packet_formats}, we summarize the different packet formats and sizes. Note that packets of type $DRp$, $BRp$ and $RRp$ are really sequences of micro-frames. The two first bits of the $Seq$ field are used to differentiate the micro-frames of a $DRp$ (00), $BRp$ (01) and $RRp$ (10); the remaining 6 are used to indicate the number of remaining micro-frames. As discussed above, the $0xX222$ and address fields are a legacy of the 802.15.4. As a consequence, the destination address is always set to 0xffff, the broadcast address. We use the source address to identify the sender.

In the payload of the $DRp$ and $BRp$ micro-frames, we specify the address of the source node (as we have less than 256 nodes in our network, 1 byte is enough to identify each node). The payload of the $RRp$ micro-frames is not used. Similarly, the $Seq$ field of the $ACK$ and $DATA$ messages is not used. The payload of the $DATA$ messages consists of two parts: the first one contains the sequence of traversed nodes needed by the routing protocol, the second the list of neighbors of the source node. Both fields are identified using a 1 byte length field in the data payload. The number of node addresses in the sequence and neighbor list must total up to less than 119.

\begin{table*}
\begin{center}
\begin{tabular}{|p{0.25in}|p{0.25in}|p{0.25in}|p{0.25in}|p{0.25in}|p{0.25in}|p{0.25in}|p{0.25in}|p{0.25in}|p{0.25in}|}
\hline
\multicolumn{2}{|c|}{0x6222} & \multicolumn{1}{|c|}{Seq.}& \multicolumn{2}{|c|}{destination addr.} & \multicolumn{2}{|c|}{source addr.} & \multicolumn{1}{|c|}{payload} & \multicolumn{2}{|c|}{Check Seq.}\\
\hline
& & & & & & & & & \\
\end{tabular}
\vspace{1mm}
\\micro-frame
\vspace{3mm}

\begin{tabular}{|p{0.25in}|p{0.25in}|p{0.25in}|p{0.25in}|p{0.25in}|p{0.25in}|p{0.25in}|p{0.25in}|p{0.25in}|}
\hline
\multicolumn{2}{|c|}{0x4222} & \multicolumn{1}{|c|}{Seq.}& \multicolumn{2}{|c|}{destination addr.} & \multicolumn{2}{|c|}{source addr.} & \multicolumn{2}{|c|}{Check Seq.}\\
\hline
& & & & & & & & \\
\end{tabular}
\vspace{1mm}
\\ACK
\vspace{3mm}

\begin{tabular}{|p{0.25in}|p{0.25in}|p{0.25in}|p{0.25in}|p{0.25in}|p{0.25in}|p{0.25in}|p{0.25in}|p{0.25in}|p{0.25in}|p{0.25in}|p{0.25in}|}
\hline
\multicolumn{2}{|c|}{0x2222} & \multicolumn{1}{|c|}{Seq.}& \multicolumn{2}{|c|}{destination addr.} & \multicolumn{2}{|c|}{source addr.} & \multicolumn{3}{|c|}{payload} & \multicolumn{2}{|c|}{Check Seq.}\\
\hline
& & & & & & & & \multicolumn{1}{|c|}{$\cdots$} & & & \\
\end{tabular}
\vspace{1mm}
\\DATA
\vspace{3mm}

\begin{tabular}{|p{0.25in}|p{0.25in}|p{0.25in}|p{0.25in}|p{0.25in}|p{0.25in}|p{0.25in}|p{0.25in}|p{0.25in}|p{0.25in}|p{0.25in}|p{0.25in}|}
\hline
\multicolumn{1}{|c|}{length} & \multicolumn{6}{|c|}{sequence of traversed nodes} & \multicolumn{5}{|c|}{neighbor list}\\
\hline
& & & & & & & & & & & \\
\end{tabular}
\vspace{1mm}
\\DATA payload
\vspace{3mm}

\caption{Packet formats at MAC level. Each graduation represents one byte.}
\label{table:packet_formats}
\end{center}
\end{table*}

Table~\ref{table:durations} summarizes the durations of the different packets and timers. For a generic packet $X$, its duration is identified by $D_X$, $T_X$ for its period (if applicable), $B_X$ for the backoff used when sending it (if applicable). Note that backoff $B_X$ is drawn within the contention window $W_X$. When a packet is identified by $Xp$, it means it is a sequence of micro-frames ($p$ stands for preamble). Note that the calculation of $W_{BR}$ and $W_{RR}$ are explained in subsection~\ref{collision}.

\begin{table}
\begin{center}
\begin{tabular}{|l|l|l|}
\hline
$D_{mf}$ & $512 µs$ & 8 bits Seq + 8 bits payload \\
\hline
$T_{mf}$ & $930 µs$ & \\
\hline
$D_{cca}$ & $1442 µs$ & $D_{mf}+T_{mf}$\\
\hline
$T_{cca}$ & $140 ms$ & $100 \times D_{cca}$ to have 1\% idle radio use\\
\hline
\hline
$D_{ACK}$ & $480 µs$ & 8 bits Seq\\
\hline
$D_{DATA}$ & $4 ms$ & 128-8=120 data bytes\\
\hline
\hline
$D_{DRp}$ & $144 ms$ & 155 micro-frames\\
\hline
$T_{DR}$ & $200 ms$ & $>D_{DATA}+D_{DRp}$\\
\hline
\hline
$T_{BRp}$ & $300 ms$ & $>2\cdot D_{BRp}+W_{BR}$\\
\hline
$D_{BRp}$ & $144 ms$ & 155 micro-frames\\
\hline
$W_{BR}$ & $10 ms$ & less than 10\% collision probability\\
\hline
$B_{BR}$ &  \multicolumn{2}{l|}{randomly and uniformly chosen in $[0..W_{BR}]$}\\
\hline
\hline
$B_{SRC}$ & $1000 ms$ & $>6(D_{BRp}+W_{BR})$\\
\hline
$D_{RRp}$ & $144 ms$ & 155 micro-frames\\
\hline
$W_{RR}$ & $30 ms$ & less than 10\% collision probability\\
\hline
$B_{ACK}$ & \multicolumn{2}{l|}{proportional to metric (uniformly distributed)}\\
\hline
$B_{RR}$ & $500µs$ & > 0\\
\hline
\end{tabular}
\caption{Timers and durations}
\label{table:durations}
\end{center}
\end{table}


\subsection{Achievable communication ranges}
\label{ranges}

During the early stages of the project, we have performed some communication range measurements using the EM2420 nodes. Results are presented in Table~\ref{table:range}. These measurements showed that the height of a node has a significant impact on the transmission range. In order to have a small network (in large networks, people tend to leave nodes behind during experimentation), we have decided to use a fixed transmission power of -25dBm.

\begin{table}
\begin{center}
\begin{tabular}{|l|r|r|}
\hline
transmission power & height & range \\
\hline
0 dBm & 1 m & 100 m \\
\hline
-25 dBm & 1 m & 25 m \\
\hline
-25 dBm & 0 m & 5 m \\
\hline
\end{tabular}
\caption{Range}
\label{table:range}
\end{center}
\end{table}


\section{Real-time verification}
\label{real-time}

Real-time systems can be divided in two classes. Hard real-time systems guarantee that a certain event happens before a given deadline. Guaranteeing involves some form of formal validation. Due to the hazardous nature of the wireless medium, and the unreliability of sensors nodes, hard-real time communication protocols for wireless sensor networks are often based on unrealistic assumptions such as a Unit Disk Graph propagation model. Soft real-time systems are made so that a portion of events happens within time bounds.

Because of link unreliability, the random nature of deployment and the path followed by the $MS$, our communication architecture can not guarantee hard-real time constraints. Rather than a hard-real time validation (based on formal models and static parameters), in this section we use mathematical models to show real-time constraints are validated in bad-case scenarios.

The critical parameter when considering real-time in our setting is the speed of the $MS$. Indeed, the network needs to broadcast the request and return the answer before the $MS$ leaves the network. The calculations presented in \ref{talking to BS} and \ref{talking to network} aim at finding a maximum speed $v_{max}$ for the $MS$.


\subsection{Goals and assumptions}

\begin{figure}
\centering
\includegraphics[width=0.50\textwidth]{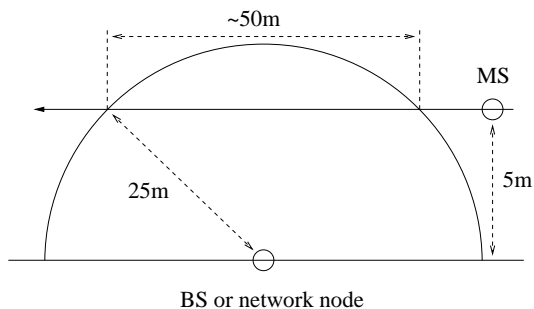}
\caption{Maximum distance over which the $MS$ can travel while connected to the $BS$ or a network node.}
\label{figure:communication_range}
\end{figure}

We assume the $MS$ moves at an altitude of 5m above the nodes. Moreover, as we use a transmission power of -25 dBm, we consider that the network and $BS$ can communicate with the $MS$ for up to 25m, according to \ref{ranges}. As depicted in Fig.~\ref{figure:communication_range}, the $MS$ is connected to another node for a duration corresponding to a movement of 50m.

\subsection{Communication between the $MS$ and the $BS$}
\label{talking to BS}

Communication between $MS$ and $BS$ goes on in phases 1 and 6. In this analysis, we consider only phase 6 which is the worst case with $DATA$ being a longer message than $ACK$. In this case we have

\begin{displaymath}
\begin{array}{lll}
v_{max} & = & \frac{50}{T_{DR}+D_{DRp}+D_{DATAmax}}\\
        & \approx & 500 km.h^{-1}
\end{array}
\end{displaymath}

We argue that this requirement is not hard to meet as, to our knowledge, no radio controlled plane achieves such speeds.


\subsection{Communication between the $MS$ and the network}
\label{talking to network}

This problem is somehow more complex than the previous one because (1) the $MS$ communicates with the complete network rather than with an individual node and (2) communication inside the network is complex and consists of broadcasting the request and transmitting the reply. We make the following assumptions.

\begin{figure}
\centering
\includegraphics[width=0.30\textwidth]{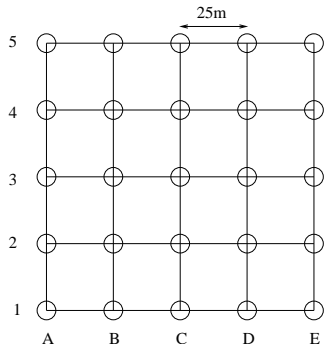}
\caption{The regular 25 node grid used to calculate $v_{max}$. The average number of neighbor nodes $N=3.20$.}
\label{figure:5x5grid}
\end{figure}

The network consists of 25 nodes regularly deployed in a square grid of size 5 hops, as depicted in Fig.~\ref{figure:5x5grid}. Two nodes on the same horizontal or vertical line are separated by 25m. Using the simulation framework defined in Section~\ref{simulation}, we obtain that the average number of hops using our communication scheme is 8.771 with a 95\% confidence interval $[8.558 \ldots 8.984]$. As a consequence, evaluating the worst case hop count at 10 is a reasonable choice. We assume the $MS$ will traverse the network entering one side, and leaving at the opposite side. The distance during which the $MS$ is connected to the network is thus between 150m and 190m.

The broadcasting protocol is blind flooding. Using the simple protocol described in Section~\ref{setup}, each node will send one $BRp$, and two neighbor nodes can not send at the same time. With the regular grid topology, the broadcast message will take up to a duration of $8\cdot\left(W_{BR}+D_{BRp}\right)$ to reach the source node, \textit{i.e.} when the node initiating the broadcast and the source in opposite corners. $B_{SRC}$ needs to be set so that the first $RRp$ message does not collide with a remaining $BRp$, \textit{i.e.} the broadcast storm needs to be over. We assume that a node has at most 6 neighbors, which is conservative considering our topology. In the worst case scenario, all these neighbors hear one another, and each has a $BRp$ message to send. Sending these messages will take at most $6\cdot\left(W_{BR}+D_{BRp}\right)$. As a consequence $B_{SRC}>6\cdot\left(W_{BR}+D_{BRp}\right)$.

Using these observations, we obtain :

\begin{displaymath}
\begin{array}{lll}
v_{max} & = & \frac{150}{8\cdot\left(W_{BR}+D_{BRp}\right)+B_{SRC}+10\cdot\left(D_{RRp}+W_{RR}+D_{DATA}\right)}\\
        & \approx & 140 km.h^{-1}
\end{array}
\end{displaymath}

We argue that this value is reasonable for a $MS$ mounted on a radio-controlled plane, which flies at a speed of about $50 km.h^{-1}$. For more demanding applications where the mobile sink is expected to go faster, it is possible to reduce $T_{mf}$, thus the time between successive clear channel assessments. This would enable the preamble messages to be arbitrarily short in time, thus speeding up the multi-hop communication, thus increasing $v_{max}$. Yet communication speed trades off with energy consumption, and reducing $T_{mf}$ increases the idle radio use. This is particularly constraining when the network sits idle most of the time.


\section{Simulation results}
\label{simulation}


\subsection{Collision probability at MAC level}
\label{collision}

We have used joint analysis and simulation to determine the collision probability between messages. Collision can happen at two instants (1) during the broadcast of a message in phases 2 and 3, denoted $P(BR)$ and (2) between the $ACK$ message during the routing procedure in phases 4 and 5, called $P(RR)$.

We calculate these collision probabilities for an average number of neighbors $N=5$. We chose to use a number larger than the average value for the regular deployment in Fig.~\ref{figure:5x5grid}, to have a security margin as the collision probability increases with the number of neighbor nodes.

\textbf{Calculating $P(BR)$.} We use the following assumptions. When a node hears a $BR$ for the first time, it starts a backoff $B_{BR}$ randomly taken within the contention window $W_{BR}$. During this duration, it remains in reception state to detect a possible relay of the message by another node. If this has not happened when its backoff timer elapses, it switches to transmission mode and relays the $BR$. There is a possibility of collision because switching from reception to transmission mode takes $D_{RxTx}=192µs$.

An analogy can be drawn between this collision probability and the one calculated in \cite{watteyne07reducing}. In this work, the authors calculate the collision probability which involved the first $ACK$ message. Collision was defined as two messages overlapping in time. Here, we can use the exact same definition, only collision is defined as another node picking a backoff time shorter than $D_{RxTx}$ after the first backoff timer expires, which is strictly equivalent to the calculation done in \cite{watteyne07reducing} but considering messages of duration $D_{RxTx}$. The theoretical value of $P(BR)$ is given in Eq.~\ref{eq:p_br}. We are not surprised to see that $P(BR)$ decreases with $W_{BR}$ increasing.

\begin{equation}
P_{BR}=1-\left(\frac{W_{RR}-D_{RxTx}}{W_{RR}}\right)^N. 
\label{eq:p_br}
\end{equation}

\textbf{Calculating $P(RR)$.} We use the results from \cite{watteyne07reducing} to calculate $P_{RR}$ in Eq.~\ref{eq:p_dr}. Similarly as in the previous case, $P(RR)$ decreases with $W_{RR}$ increasing.

\begin{equation}
P_{RR}=1-\left(\frac{W_{RR}-D_{ACK}}{W_{RR}}\right)^N. 
\label{eq:p_dr}
\end{equation}

\begin{figure}
\centering
\includegraphics[width=0.70\textwidth]{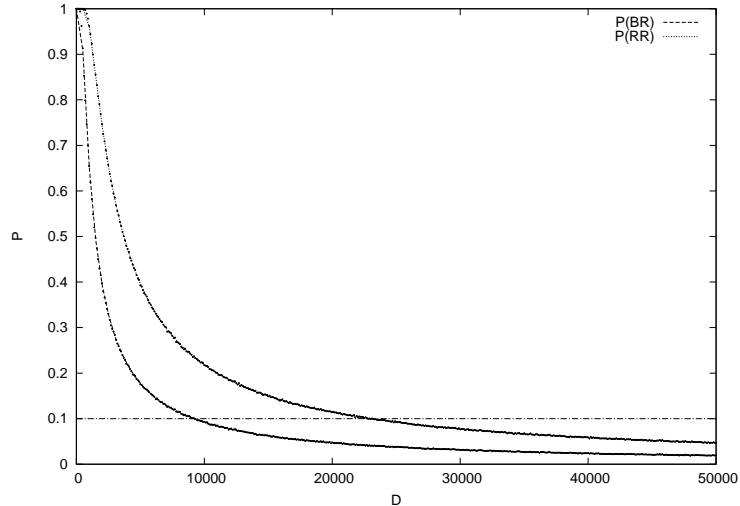}
\caption{Collision probability for $N=5$. The theoretical results are presented as plain lines whereas the results obtained by simulation are represented as unconnected dots. To ease readability, we have plotted $P=0.1$ . Simulation results are averaged over $10^5$ runs.}
\label{figure:p_vs_d}
\end{figure}

We want the collision probability in either cases to be lower than 10\%, which we consider an acceptable collision rate. To derive the value of both contention windows $W_{BR}$ and $W_{RR}$, we plot Fig.~\ref{figure:p_vs_d} using $D_{ACK}$ from Table~\ref{table:durations}. The simulation results provided by an iterating C++ program are presented as dots in Fig.~\ref{figure:p_vs_d} and match the theoretical results. We see that with $W_{RR}=30ms$ and $W_{BR}=10ms$ we achieve $P<0.1$.

\subsection{Routing protocol on a random graph}
\label{routing_random}

Evaluating the performance of a routing protocol is a task typically done by simulation, as routing can be seen as a complex global behavior emerging from simple local interactions between nodes. We ran these simulations on a home-made C++ simulator. In this subsection, we chose to use a graph where nodes are randomly positioned, the $X$ and $Y$ coordinates being randomly positioned within $[0 \ldots 1000]$. We assume a constant communication range of 200, and a simple Unit Disk Graph propagation model. We vary the average number of neighbor nodes, and calculate the number of nodes accordingly. Each node runs a perfect MAC protocol (which does not model collisions) and the routing protocol described in subsection~\ref{adapting_3rule}. Each message is sent from a randomly and uniformly chosen node (connected to the sink) to the sink node.

\begin{figure}
\centering
\includegraphics[width=0.70\textwidth]{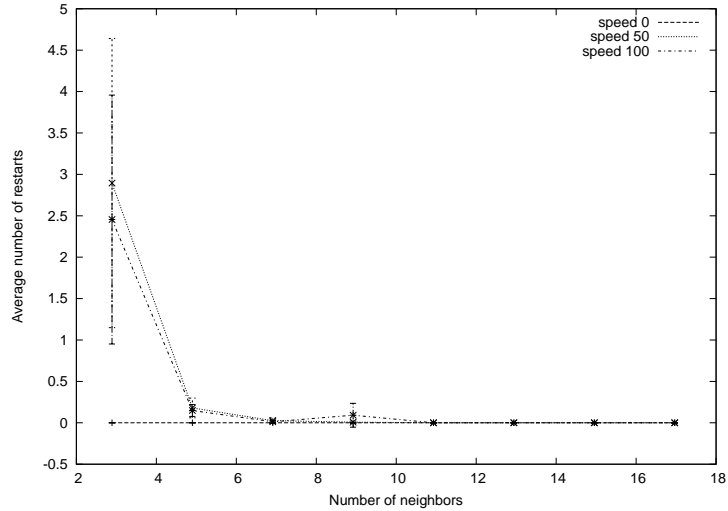}
\caption{Number of restarts on a \textbf{random graph}. The number of restarts is relatively low, and decreases with sink speed decreasing and number of neighbors increasing. A 95\% confidence interval is presented with the data.}
\label{figure:restarts}
\end{figure}

\begin{figure}
\centering
\includegraphics[width=0.70\textwidth]{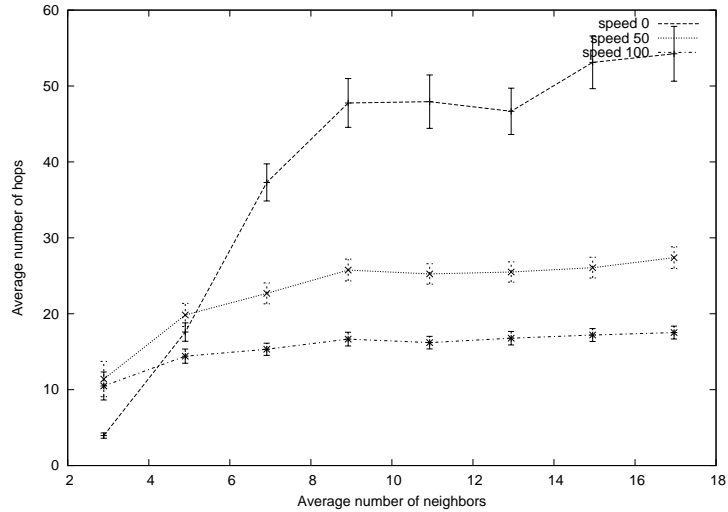}
\caption{Hop count on a \textbf{random graph}. Whereas sink node mobility increases the number of restarts, thus the number of potential hops, it globally decreases the number of hops. A 95\% confidence interval is presented with the data.}
\label{figure:results}
\end{figure}

Simulation is performed in rounds. At each round, a node decides which of its neighbors is the next hop according to our routing protocol, and sends its data. At the same time, we update the sink node's position as follows. The sink node's $X$ position is increased by a number chosen randomly and uniformly between 0 and a maximum value (called $speed$ in Fig.~\ref{figure:restarts} and Fig.~\ref{figure:results}). When the $X$ position reaches the border of the field (here 1000), it is decreased (at each iteration) until it reaches zero. The same algorithm is applied to the sink nodes $Y$ category. As a results, the sink node moves in a direction picked in [north-east, south-east, south-west, north-west], and bounces off the edges of the field much like a ball on a pool table.

The first result we want to extract is the number of restarts the routing protocol undergoes. Although this feature makes the protocol robust to link dynamics and sink movement, we want to keep the number of restarts low as it increases the number of hops. We depict the number of restarts versus the average number of neighbors in Fig.~\ref{figure:restarts}. The number of restarts is low in all runs. Note that the 95\% confidence interval looks large because of the low values of the number of restarts. Yet, we see that the number of restarts decreases when the number of neighbors increases and the sink speed decreases. A first recommendation would be to keep the speed of the $MS$ as low as possible. We will see in the next paragraph that this is not necessarily true.

In Fig.~\ref{figure:results}, we plot the number of hops for a message to reach the sink node. The fact that this number increases with the number of neighbors should not be misunderstood. Indeed, with a low average number of neighbors, the source node is necessarily close to the sink, as the probability for a node to be connected decreases quickly with distance\cite{onat07generating}. The surprising results here is that the number of hops decreases when sink mobility increases. This is somehow contradictory with the previous observation, as sink mobility increases the number of restarts, thus hop count. Yet, sink mobility increases the probability that the sink encounters the message during its transmission. Fig.~\ref{figure:results} shows that this behavior outbalances the increased hop count due to routing protocol restarts.

\subsection{Routing protocol on the regular graph}

\begin{figure}
\centering
\includegraphics[width=0.70\textwidth]{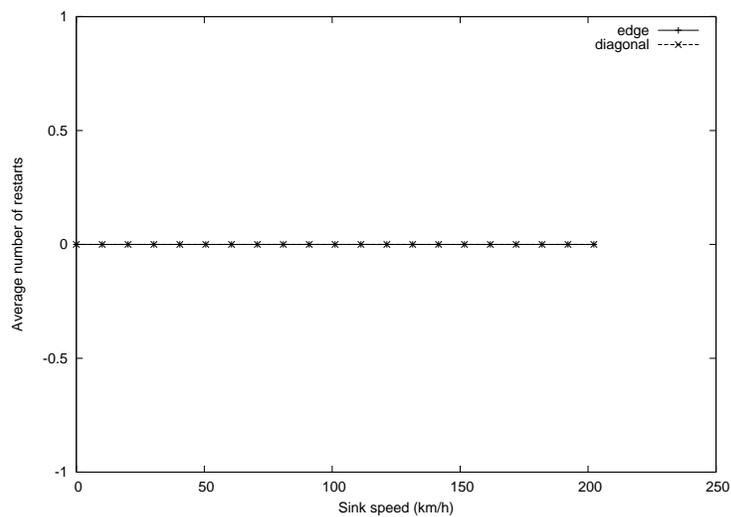}
\caption{Number of restarts on a \textbf{regular graph}.}
\label{figure:restarts25}
\end{figure}

\begin{figure}
\centering
\includegraphics[width=0.70\textwidth]{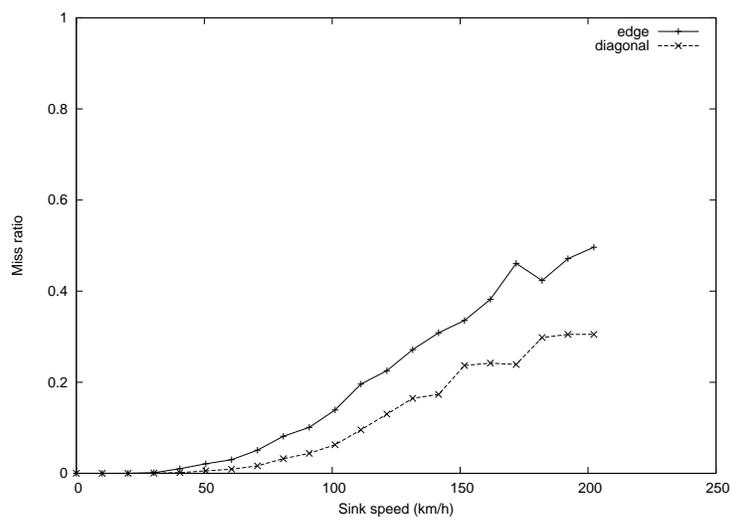}
\caption{Ratio of missed messages because of network time-out on a \textbf{regular graph}. Simulation results are averaged over $10^4$ runs.}
\label{figure:missed25.eps}
\end{figure}

\begin{figure}
\centering
\includegraphics[width=0.70\textwidth]{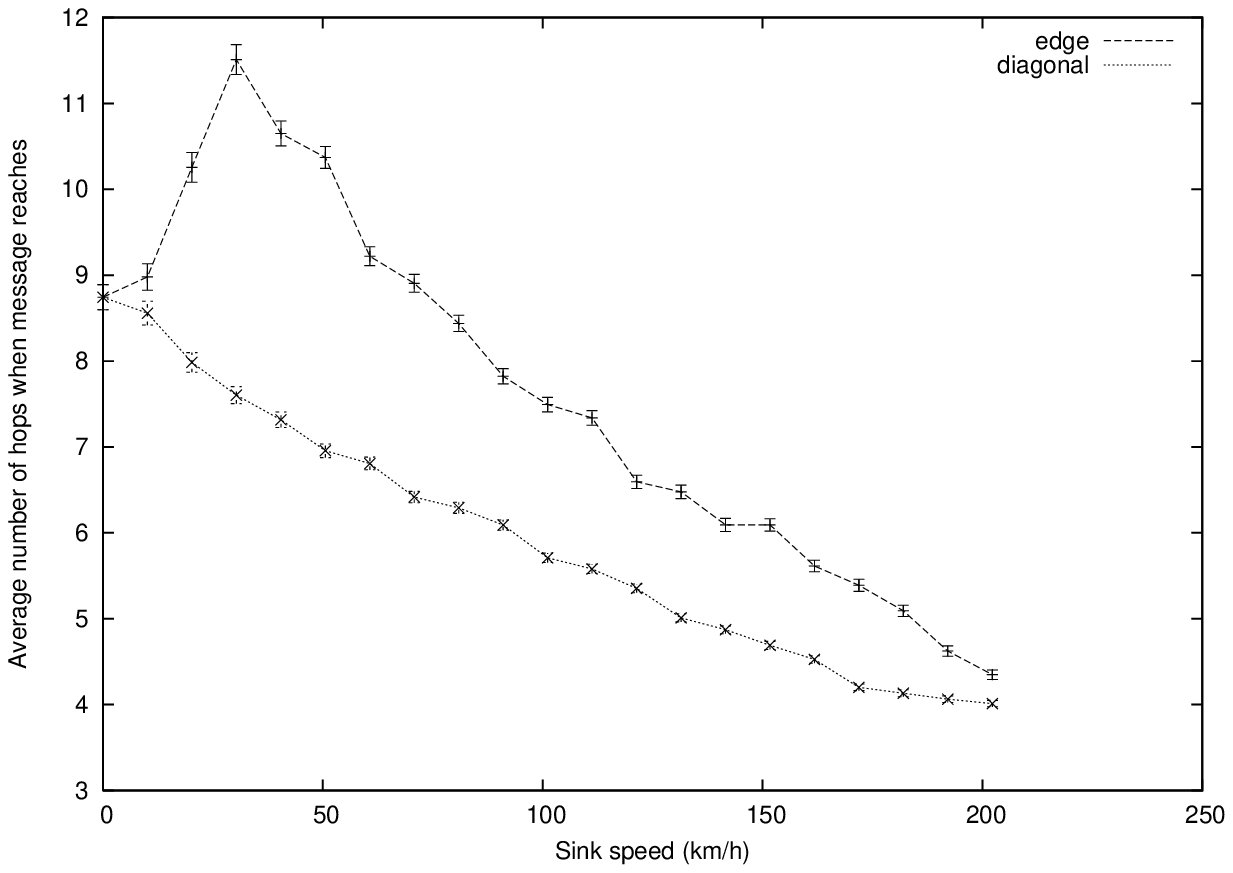}
\caption{Hop count on a \textbf{regular graph}. Simulation results are averaged over $10^4$ runs.}
\label{figure:results25}
\end{figure}

Subsection~\ref{routing_random} presents results on a random graph, with a random sink movement. Moreover, the sink always stays connected to the network. Yet, we would like to perform similar simulations on the regular graph of Fig.~\ref{figure:5x5grid} to have data comparable to the experimental results. Our topology is thus a regular grid of 25 nodes. The total network is a square of size 100m. We assume the $MS$ leaves the network by the side opposite to the one it entered. We assume that it moves in a straight line, at a constant speed. We have two mobility models. In the first one (called $edge$ in Figs.~\ref{figure:restarts25}-\ref{figure:missed25.eps}-\ref{figure:results25}), the $MS$ enters in the lower-left corner, and leaves at the lower right corner. In the second one (called $diagonal$ in Figs.~\ref{figure:restarts25}-\ref{figure:missed25.eps}-\ref{figure:results25}), it enters at the lower-left corner but leaves at the upper-right corner. These two models represent the shortest and longest duration the $MS$ is connected to the network, respectively.

With the sink leaving the network, it is now possible that the network times-out, \textit{i.e.} the $MS$ has already left the network when the message should have reached it. We therefore can have a non-zero miss ratio, the ratio of the messages not reaching the $MS$.

In Fig.~\ref{figure:restarts25}, we plot the number of restarts as a function of the sink speed. We see that, due to the fact that the $MS$ is only connected to the network for a limited duration, the routing protocol has no time to trigger restarts.

As stated before, it is possible than the $MS$ moves too fast to allow a message to reach it. Fig.~\ref{figure:missed25.eps} depicts miss ratio against sink speed. As expected, this number increases with the sink speed. Moreover, as moving along the diagonal allows more time for the message to reach the sink, the miss ratio is less.

One should be careful when reading Fig.~\ref{figure:results25} as it shows the number of hops needed for a message to reach the sink where only messages which actually reach it are taken into account. The fact that this number decreases with the sink speed increasing has two causes: (1) the sink encounters the message and (2) for a high sink speed, the miss ratio being high, successful transmission originate from source nodes already close to the sink's trajectory.

\section{Off-site experimental results}
\label{off-site experimental}


\subsection{Energy consumption of the 1hopMAC protocol}

\begin{figure}
\centering
\includegraphics[width=0.70\textwidth]{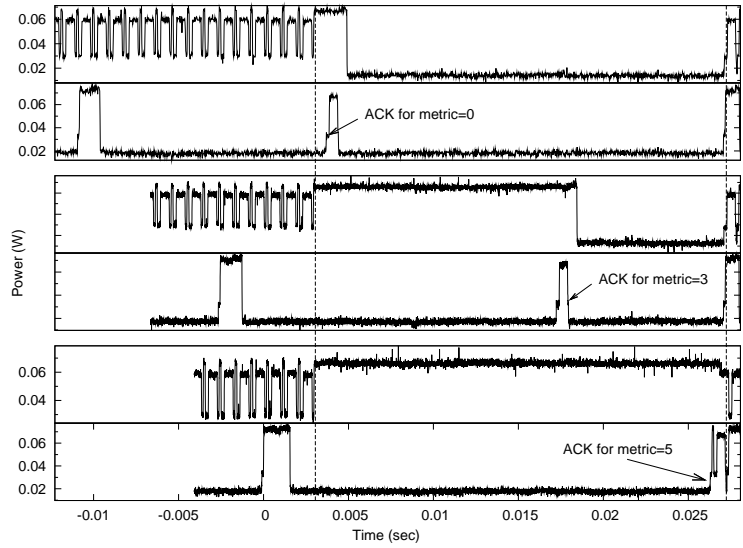}
\caption{Power consumption versus time during Phase 4 with two communicating nodes. Measurements for a transmitter and the receiver are presented in the upper and lower parts, resp. The experiment is repeated for a metric equal to 0, 3 and 5, hence the three groups of figures. Note that the first part of the preamble is truncated to ease readability.}
\label{figure:power_1hopmac}
\end{figure}

Prior to the demonstration results, we study the energy consumption of our communication stack. As it is the MAC protocol which controls the state of the radio module (transmission, reception or idle), with a given PHY layer, energy efficiency is primarily a MAC-layer issue. As a first experimental setting, we read the power consumption using a oscilloscope attached directly to the power source of nodes. We will analyze in more detail the power consumption values and different duration, but let's first focus on Fig.~\ref{figure:power_1hopmac} which plots the power consumption as a function of time at a sending (upper part) and a receiving node (lower part). We repeat this experience for different metric values.

We have drawn vertical lines to ease the interpretation of the data presented in Fig.~\ref{figure:power_1hopmac}. On the leftmost part, the sending node sends a series of microframes; the receiver wakes up and receives one micro-frame. From the information it contains, the receiver is put back to sleep until the last micro-frame is sent. In the central part, the sender starts listening for an $ACK$ message, which the receivers sends after a backoff proportional to its metric. As for these measurements we have used the first variant of 1-hopMAC (refer to \cite{watteyne061hopmac} for details), the sender switches off its radio after successfully receiving a first $ACK$. In the rightmost part, after the contention window for the $ACK$ messages has passed, both sending and receiving nodes switch their radios back on. The sender starts by informing the receiver it has been selected as the next hop, and sends the $DATA$ to it.

This simple 2-node setting provides us with interesting information:
\begin{itemize}
   \item The radio module has a major impact on the total energy budget of a node and its activity can be directly read from the power consumption of the whole node;
   \item Sending, receiving or idle listening consume approximately the same amount of energy;
   \item We verify our implementation by noting that the $ACK$ messages are sent at different instants depending on the value of the node's metric.
\end{itemize}

Now that we have a clear picture of the functioning of our implementation, we extract the energy consumption of the different radio states. We do this analysis for transmission at both 0dBm and -25dBm, and for both types of nodes. We make sure that all measured timers and packet duration are consistent with the requirements from Table~\ref{table:durations}. The first set of results is presented in Table~\ref{table:consumption_states}, and relate to the power consumption of individual states of the radio chip. Using the consumption model, we can start analyzing more macroscopic behaviors.







\begin{table}
\begin{center}
\begin{tabular}{|l|r|r|}
\cline{2-3}
\multicolumn{1}{c}{} & \multicolumn{2}{|c|}{EM2420 module} \\
\cline{2-3}
\multicolumn{1}{c|}{} & \multicolumn{1}{|c|}{0dBm} & \multicolumn{1}{|c|}{-25dBm} \\
\hline
$P_{sleep}$ & 8.018 mW & 2.735 mW \\
\hline
$P_{poll}$ & 8.629 mW & 3.300 mW \\
\hline
$P_{listen}$ & 65.833 mW & 61.030 mW \\
\hline
$P_{Tx}$ & 66.156 mW & 32.807 mW \\
\hline
$P_{Rx}$ & 70.686 mW & 65.444 mW \\
\hline
\end{tabular}
\caption{Consumption of the individual radio states}
\label{table:consumption_states}
\end{center}
\end{table}

Let's consider we are using the EM2420 module with its transmission power set to -25dBm. This module is powered by two AAA alkaline batteries, providing a total of about 10,000 J (2.8 Wh). When no activity goes on, all nodes operate in preamble sampling mode. We have measured that this mode consumes 3.300 mW on average, offering a lifetime of 850 hours, or 35 days. A MAC protocol such as IEEE802.11 which leaves the radio modules on would consume 61.030 mW during idle period, offering a lifetime of only 45 hours, even without traffic.

To complete our energy consumption model, we derive the energy needed for sending and receiving a packet. Note that relaying a packet is equivalent to receiving and retransmitting it. We consider for these calculations that each node has 5 neighbors on average. Sending a packet is equivalent to sending a preamble, then listening and receiving 5 $ACK$ messages, and finally sending the $DATA$ message. This results in energy consumption of $E_{Tx}$. Among the 5 neighbors, all will consume the energy equivalent to receiving one preamble, and sending an $ACK$ ($E_{Comp}$). Among the 5 neighbors, only one will have the additional cost of receiving the data, resulting in a total energy expenditure of $E_{Rx}$. Thus, sending a packet under these assumptions costs $E_{Tx}+E_{comp}+E_{Rx}$. Refer to Eq.~\ref{eq:consumption_phases} for the detailed calculation.

\begin{equation}
\begin{array}{ll}
E_{Tx}   = & E_{preamb}+(W_{RR}-N \cdot D_{ACK})P_{listen}+\\
           & (N \cdot D_{ACK})P_{Rx}+D_{DATA}P_{Tx}\\
E_{comp} = & ({D_{RRp}+W_{RR}}+D_{DATA}-D_{cca}-D_{ACK})\times\\
           & P_{sleep}+P_{Rx}D_{cca}+P_{Tx}D_{ACK}\\
E_{Rx}   = & ({D_{RRp}+W_{RR}}-D_{cca}-D_{ACK})P_{sleep}\\
           & +P_{Rx}D_{cca}+P_{Tx}D_{ACK}+P_{Rx}D_{DATA}\\
\end{array}
\label{eq:consumption_phases}
\end{equation}

\begin{table}
\begin{center}
\begin{minipage}{0.70\textwidth}
\begin{tabular}{|l|r|r|}
\cline{2-3}
\multicolumn{1}{c}{} & \multicolumn{2}{|c|}{EM2420 module} \\
\cline{2-3}
\multicolumn{1}{c|}{} & \multicolumn{1}{|c|}{0dBm} & \multicolumn{1}{|c|}{-25dBm} \\
\hline
$E_{preamb}$\footnote{This value is measured, the others are derived from Tables.~\ref{table:durations} and \ref{table:consumption_states}.} & 1.243 mJ & 0.467 mJ \\
\hline
$E_{Tx}$ & 3.494 mJ & 2.440 mJ \\
\hline
$E_{comp}$ & 1.545 mJ & 0.592 mJ \\
\hline
$E_{Rx}$ & 1.796 mJ & 0.843 mJ \\
\hline
\end{tabular}
\end{minipage}
\caption{Consumption of major protocol phases using $1-hopMAC_{basic}$, with a mean number of neighbors of 5}
\label{table:consumption_phases}
\end{center}
\end{table}

Table~\ref{table:consumption_phases} provides us with some interesting insights on the functioning of our MAC protocol. First, all nodes are virtually on at any time, avoiding costly (re-)synchronization mechanisms. Second, it enables continuous tuning of the latency/energy efficiency trade-off through tuning $T_{mf}$. Finally, we see that our protocol transfers most of the energy burden of communication to the sender. Indeed, sending a message costs about 2-3 times more energy than receiving one. This has a major impact on upper-layer protocol design as having a dense network does not jeopardize energy-efficiency.

By coupling the 1-hopMAC protocol with the 3rule routing protocol and virtual coordinates, we self-organize the network in a very cost effective way as no structure needs to be maintained when the network sits idle.


\subsection{Collision probability at MAC level}

The aim of the 1-hopMAC protocol is to avoid maintaining a neighborhood table. Instead, this table is implicitly built on-demand with nodes acknowledging a request during a contention window. All nodes send their $ACK$ message after a backoff proportional to some metric, which we consider uniformly distributed in some range. Several $ACK$ messages may collide, in which can these messages are lost. As the requesting node choses the node which answers first as the next hop, a collision involving this first message can result in choosing the wrong next hop node. This phenomenon has been analyzed in \cite{watteyne07reducing}, with the simple assumption that all messages overlapping in time collide and are lost (see subsection~\ref{collision}). Yet, this assumption does not take into account the capture effect which happens when a receiving node successfully decodes at least one of the messages which were overlapping. In other words, the assumption used in \cite{watteyne07reducing} may be too pessimistic. The aim of this subsection is to verify and quantify this.

\begin{figure}
\centering
\includegraphics[width=0.50\textwidth]{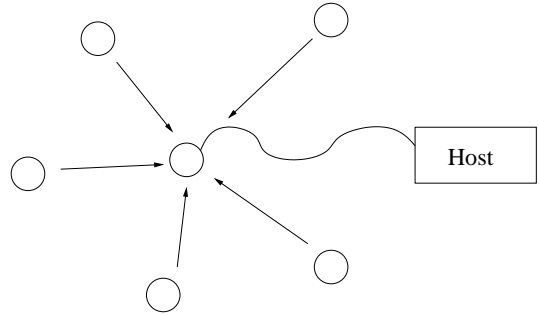}
\caption{The experimental setting used to determine the collision probability.}
\label{figure:setting_collision}
\end{figure}

The experimental setting goes as follows. As depicted in Fig.~\ref{figure:setting_collision}, a central node is connected to a host computer through the development kit provided with the EM2420 modules, with 5 battery-powered nodes surrounding it. The basic experiment is divided in two phases. During the first phase, the host computer randomly choses 5 integer metrics within $[0..361]$, and transfers them to the central node. We choose the range $[0..361]$ because this is the range the built-in random generator of the EM2420 modules operates in. The central node then broadcasts a packet containing these metrics. Each surrounding node picks one of these metrics using a predefined sequence (each node picks a different metric). In the second phase, the central node uses the 1-hopMAC protocol and issues a request. Based on the received $ACK$ messages, its takes a decision of which is the next hop. This decision is transfered back to the host computer, which compares it with the theoretical decision (\textit{i.e.} the node with lowest metric should have been chosen). By repeating this experiment a large number of times, it is possible to extract the probability 1-hopMAC takes the wrong decision, caused by collision between $ACK$ messages.

\begin{figure}
\centering
\includegraphics[width=0.70\textwidth]{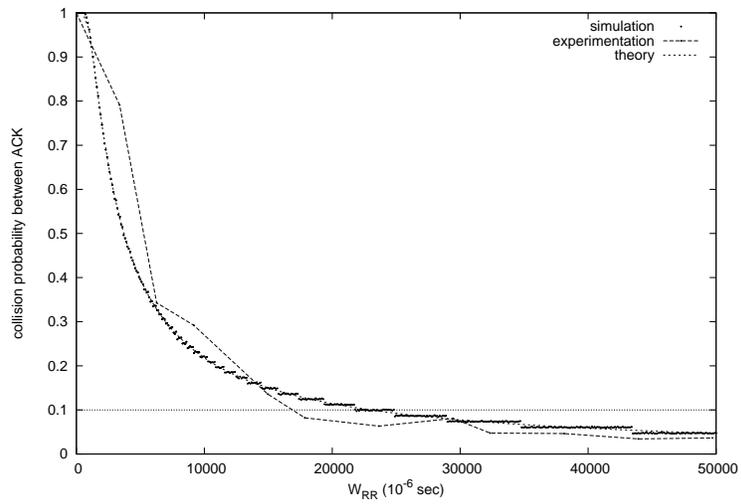}
\caption{Comparing the experimental collision probability with theoretical and simulation results. Simulation results are averaged over $10^5$; experimental results over 8000 runs.}
\label{figure:collision_prob}
\end{figure}

We plot our experimental results together with simulation and theoretical results in Fig.~\ref{figure:collision_prob}. Each node contains a metric between 0 and 361. The difference between Fig.~\ref{figure:collision_prob} and Fig.~\ref{figure:p_vs_d} is that the former was drawn with a discrete set of 362 possible backoff values, uniformly distributed in $[0 \ldots W_{RR}]$

The results confirm our analysis that the capture effect lowers the collision probability, {i.e.} simulation and analysis are too pessimistic compared to experimentation. To quantify the gain, we integrate the curves and find a 1.21\% decrease of the experimental collision probability compared to the theoretical one. As a consequence, the values of the contention window $W_{RR}$ determined in Section~\ref{collision} are valid and even a little conservative. The same result applies to $W_{BR}$.


\section{On-site experimentation results}
\label{on-site experimental}

The experimentat was carried out during the summer at Alpe d'Huez, a ski resort in the French Alps. The aircraft is a MS2001 with a wingspan of 2.20 meters, powered by a 7.5cc motor, and controlled by a Multiplex radio. It flies at about 25 km/h. During the experiments, we asked the pilots to fly above the network and the base station at an altitude of no more than 5 m.

\begin{figure}
\centering
\includegraphics[width=0.50\textwidth]{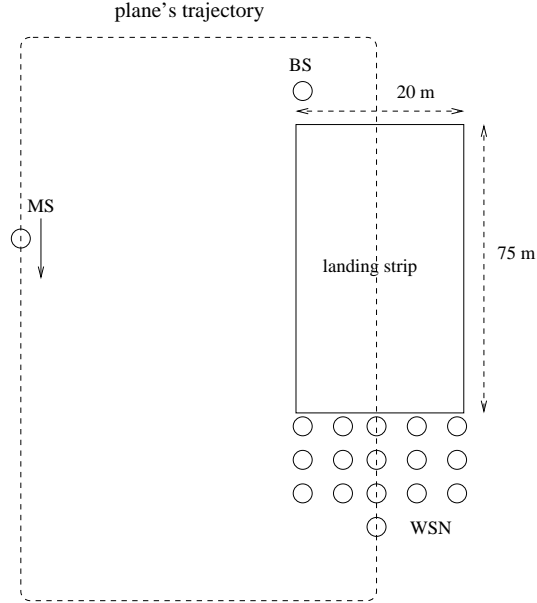}
\caption{Experimental setup of the WiFly demo (bird's view).}
\label{figure:experimental_setup}
\end{figure}

For the network and the base station to be disconnected, they are deployed at the opposite ends of the runway, about 80 m apart. As depicted in Fig.~\ref{figure:experimental_setup}, we used a 16-node network. Due to the nature of the terrain, the deployed network slightly differs from the one used in the previous Sections of this report.

\begin{figure}
\centering
\includegraphics[width=0.70\textwidth]{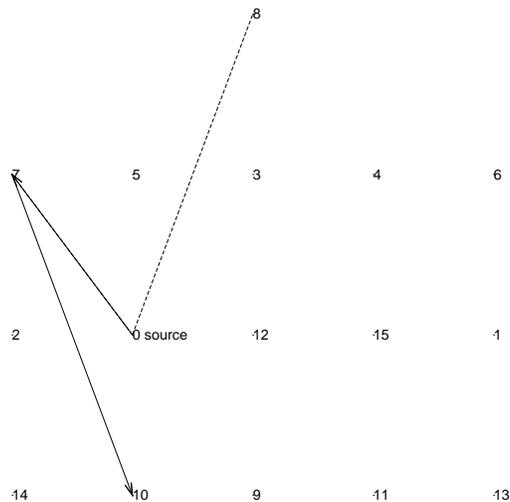}
\caption{Snapshot of the real-time display during the experiment. Each number identifies a node.  Note that the topology depicted here is upside-down compared to Fig.~\ref{figure:experimental_setup}. Dashed lines connect the source node (here 0) with its neighbors (here 7 and 8). Plain arrows indicate the path between the source node and the MS (here 0->7->10).}
\label{figure:display}
\end{figure}

After some adjustments, the experiment ran smoothly. A laptop was connected to the $BS$ and drew the neighbors of the source node as well as the path followed by the last message, in real time. Fig.~\ref{figure:display} shows such a snapshot. It is interesting to note that due to the random nature of the electromagnetic signal, neighbors nodes are not always the closest ones (\textit{e.g.} node 12 is not node 0's neighbor in Fig.~\ref{figure:display}). We refer the interested reader to the first author's website for complementary photos and videos of the experiment.

\section{Conclusion and future work}
\label{conclusion}

In this report, we have presented a complete energy-efficient self-organizing communication architecture, composed of the 1-hopMAC and the 3rule routing protocols, used over virtual coordinates. The application involves a mobile sink going back and forth between a base station which issues requests, and a wireless sensor network.

We have used simulation to show the collision probability at MAC layer, and the routing performances. These characteristics were verified by experimentation, together with the energy consumption of our platform. This platform was used in a real-world deployment, with the mobile sink mounted on a radio controlled airplane.

From a protocol point of view, current and future work involves optimizing the routing protocol by updating the virtual coordinates to have a lower hop count. We want to reuse the platform presented here, and we are currently working on documenting the different steps needed to implement a given protocol.

\section*{Acknowledgements}

The authors would like to thank Michaël Gauthier for his tremendous work and motivation for implementing our protocols, Julien Gaillard for his great work on the hardware of the mote, Tahar Jarboui for his help on Think and Loïc Amadu and Loris Grasset from "Sejours Vacances Modélisme" at Alpe d'Huez for letting us use their radio controlled planes.

\bibliographystyle{IEEEtran}
\bibliography{watteyne08wifly}

\end{document}